\documentclass[twocolumn,preprintnumbers,amsmath,amssymb,prb,superscriptaddress]{revtex4}

\usepackage{graphicx}
\usepackage{dcolumn}
\usepackage{bm}
\usepackage{amssymb}
\usepackage{epstopdf}
\usepackage{color}
\usepackage{dsfont}
\usepackage{float}

\begin{document}

\title{Optically induced topological states on the surface of mercury telluride}

\author{O. Kyriienko}
\affiliation{NORDITA, KTH Royal Institute of Technology and
Stockholm University, Roslagstullsbacken 23, SE-106 91 Stockholm,
Sweden} \affiliation{ITMO University, St. Petersburg 197101,
Russia}

\author{O. V. Kibis}\email{Oleg.Kibis(c)nstu.ru}
\affiliation{Department of Applied and Theoretical Physics,
Novosibirsk State Technical University, Karl Marx Avenue 20,
630073 Novosibirsk, Russia} \affiliation{Science Institute,
University of Iceland IS-107, Reykjavik, Iceland}

\author{I. A. Shelykh}
\affiliation{ITMO University, St. Petersburg 197101, Russia}
\affiliation{Science Institute, University of Iceland IS-107,
Reykjavik, Iceland}


\begin{abstract}
We developed the theory which describes the Floquet engineering of
surface electronic modes in bulk  mercury telluride (HgTe) by a
circularly polarized electromagnetic field. The analysis shows
that the field results in appearance of the surface states which
arise from the mixing of conduction and valence bands of HgTe.
Their branches lie near the center of the Brillouin zone and have
the Dirac dispersion characteristic for topological states.
Besides them, the irradiation induces the gap between the
conduction and valence bands of HgTe. Thus, the irradiation can
turn mercury telluride into topological insulator from gapless
semiconductor. It is demonstrated that the optically induced
states differ substantially from the non-topological surface
states existing in HgTe without irradiation. The structure of the
found states is studied both analytically and numerically in the
broad range of their parameters.
\end{abstract}

\maketitle

\section{Introduction}
In the last years, the achievements in the laser and microwave
techniques have made possible the optical control of condensed
matter structures with a high-frequency electromagnetic field
(so-called dressing field), which is based on the Floquet theory
of periodically driven quantum systems (Floquet
engineering)~\cite{Hanggi_98,Kohler_05,Bukov_15,Holthaus_16}.
Particularly, the studies of various nanostructures strongly
coupled to light---including quantum wells
\cite{Wagner_10,Teich_13,Pervishko_15,Dini_16,Morina_15}, quantum
rings \cite{Sigurdsson_14,Kibis_13,Koshelev_15,Joibari_14},
quantum spin
chains~\cite{Russomanno_17,Potter_17,Kyriienko_18,Kennes_18},
graphene and Dirac materials
\cite{Kristinsson_16,Kibis_16,Glazov_14,Syzranov_13,Kibis_11_1,
Kibis_10,Oka_09,Kibis_17,Iurov_18,Durnev_19} etc.---have emerged
as a vibrant area of contemporary physics with the objective to
control electronic properties of these systems.

Among various low-dimensional electronic systems, electrons
localized near boundaries of condensed matter structures (surface
electronic states) bear a special role. The increasing interest of
scientific community devoted to them is caused by the
topologically nontrivial nature of the surface states in
structures known as topological
insulators~\cite{Hasan_2010,Chiu_2016,Bansil_2016} which behave
like an insulator in their bulk but have the gapless conducting
electronic modes protected by the time-reversal symmetry at their
boundaries. As a consequence, the Floquet engineering of
topological surface states (Floquet topological insulators)
attracts attention as an effective tool to control their physical
properties. Particularly, it is shown that the irradiation with a
high-frequency electromagnetic field induces topological edge
states in graphene~\cite{Perez_14,Usaj_14} and semiconductor
quantum wells~\cite{Lindner_2011,Hasan_2017}. Optically induced
Weyl points have been predicted in topological insulators and
Dirac
semimetals~\cite{Wang_2014,Zou_2016,Zhang_2016,Hubener_2017,Bucciantini_2017}.
Recently, the theory of optically-controlled spin transport on the
surface of bulk topological insulators was
elaborated~\cite{Yudin_2016}, light-induced modification of
surface topological states in thin films was
studied~\cite{Pervishko_2018}, and the optically induced
topological edge states in the array of quantum rings were
analyzed~\cite{Kozin_2018}.

In the present article, we apply the Floquet engineering approach
to the bulk gapless semiconductor---mercury telluride (HgTe).
Although this material has shown topologically nontrivial
electronic properties actively studied last years, only strain was
considered before to turn bulk HgTe into topological
insulator~\cite{Dai_2008,Brune_2011,Maier_2012,Yan_2012,Chen_2012,Kozlov_2014,Dantscher_2015,Ruan_2016}.
In contrast to this, we demonstrate theoretically that a
circularly polarized high-frequency electromagnetic field can
create topological electronic states on HgTe surface. Since the
same field opens the gap between conduction and valence bands, the
light-induced topological phase transition (which turns HgTe from
gapless semiconductor into topological insulator) occurs.

The article is organized as follows. In Sec. II, we formulate the
Hamiltonian formalism describing the electronic states on the
irradiated surface of HgTe. In Sec. III, we solve the
corresponding Schr\"odinger problem analytically in the simplest
particular cases, calculate the dispersion of the surface states
numerically, and analyze their energy spectrum. This is followed
by the conclusion.

\section{The Hamiltonian}
In the present article, we consider the surface electronic states
which are localized near the surface $(001)$ of bulk HgTe and
originate from the light-induced mixing of conduction and valence
bands near the center of the Brillouin zone. First of all, let us
write the Hamiltonian describing these bands in bulk
HgTe~\cite{Bir_Pikus} without a dressing field,
\begin{equation}\label{H}
\hat{\cal H}=\hat{\cal H}_{\mathrm{L}}+\hat{\cal
H}_{\mathrm{BIA}},
\end{equation}
where
\begin{eqnarray}\label{HL}
\hat{\cal H}_{\mathrm{L}}&=&\left(\gamma_1+{5\gamma_2}/{2}\right)
\mathbf{k}^2-2\gamma_2(J_x^2k_x^2+J_y^2k_y^2+J_z^2k_z^2)\nonumber\\
&-&2\gamma_3(\{J_x, J_y\}k_x k_y+ \{J_x, J_z\}k_x k_z+\{J_y,
J_z\}k_y k_z)\nonumber\\
\end{eqnarray}
is the Luttinger Hamiltonian,
\begin{eqnarray}\label{Hbia}
\hat{\cal H}_{\mathrm{BIA}}&=&\alpha[k_x\{J_x,
(J_y^2-J_z^2)\}+k_y\{J_y, (J_z^2-J_x^2)\}\nonumber\\
&+&k_z\{J_z, (J_x^2-J_y^2)\}]
\end{eqnarray}
is the term coming from the bulk inversion asymmetry (BIA) of the
crystal structure, $\mathbf{k}=(k_x,k_y,k_z)$ is the electron wave
vector, $\gamma_{1,2,3}$ are the Luttinger parameters, $\alpha$ is
the BIA parameter, $J_{x,y,z}$ are the $4\times4$ matrices
corresponding to the electron angular momentum $J=3/2$, and the
curly brackets $\{A,B\}$ represent the anti-commutators of the
matrices  $A$ and $B$. To perform calculations, it is convenient
to rewrite the Hamiltonian (\ref{H}) as a $4\times4$ matrix in the
basis of Luttinger-Kohn wave functions, $\psi_{j_z}$, which
describe four-fold degenerate electron states of the conduction
and valence band in the center of the bulk Brillouin zone, and
correspond to the four different projections of electron momentum
on the $z$ axis, $j_z=\pm1/2$ and $j_z=\pm3/2$ (see e.g.
Ref.~[\onlinecite{Bir_Pikus}]). In this basis, the Hamiltonian
(\ref{H}) reads
\begin{equation}\label{LHM0}
\hat{\cal H} =
\begin{tabular}{|c||c c c c|} \hline
${j_z}\backslash {j_z}$ & ${+3/2}$ & ${-1/2}$ & ${+1/2}$ & ${-3/2}$ \\
\hline\hline ${+3/2}$ & $F$ & $I+L$ & $H+M$ & $N$ \\
${-1/2}$ & $I^*+L$ & $G$ & $-N$ & $-H+M$ \\
${+1/2}$ & $H^*+M^*$ & $-N^*$ & $G$ & $I-L$ \\
${-3/2}$ & $N^*$ & $-H^*+M^*$ & $I^*-L$ & $F$ \\
\hline
\end{tabular}\,,
\end{equation}
where the matrix elements are
\begin{align}\label{LP0}
&F=(\gamma_1+\gamma_2)(k_x^2+k_y^2)+(\gamma_1-2\gamma_2)k_z^2,\nonumber\\
&G=(\gamma_1-\gamma_2)(k_x^2+k_y^2)+(\gamma_1+2\gamma_2)k_z^2,\nonumber\\
&I=-\sqrt{3}\gamma_2(k_x^2-k_y^2)+i2\sqrt{3}\gamma_3k_xk_y,\,\,L=\sqrt{3}\alpha k_z,\nonumber\\
&M=-(\sqrt{3}\alpha/2)(k_x+ik_y),\,\,H=-2\sqrt{3}\gamma_3(k_x-ik_y)k_z,\nonumber\\
&N=-(3\alpha/2)(k_x-ik_y),\nonumber
\end{align}
\begin{figure}[t]
\includegraphics[width=1\linewidth]{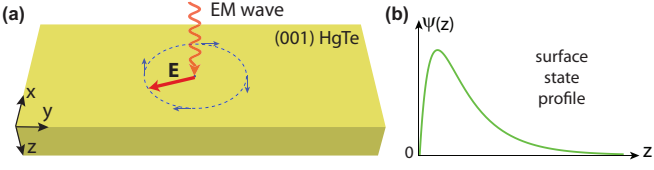}
\caption{Sketch of the system under consideration: (a) $(001)$
surface of bulk HgTe, irradiated by a circularly polarized
electromagnetic wave with the electric field amplitude
$\mathbf{E}$; (b) wave function, $\Psi(z)$, of surface electronic
states localized near the $(001)$ surface of HgTe.} \label{Fig1}
\end{figure}
Next, we add a dressing field which introduces the optically
induced mixing of conduction and valence bands. Within the
conventional minimal coupling approach, we perform the replacement
$\mathbf{k}\rightarrow e\mathbf{A}(t)/\hbar$ in the Hamiltonian
(\ref{H}), where $\mathbf{A}=(A_x,A_y,A_z)$ is the time-dependent
vector potential of the dressing field near the irradiated
surface. Assuming the electromagnetic wave (dressing field) to be
circularly polarized and propagating along the $z$ axis (see
Fig.~\ref{Fig1}a), the vector potential near the irradiated
$(001)$ surface of HgTe can be written as
\begin{eqnarray}\label{A}
\mathbf{A}=\frac{E}{\omega}(\cos\omega t,\,\sin\omega t,\,0),
\end{eqnarray}%
where $E$ and $\omega$ are the amplitude and frequency of the EM
wave, respectively. Then the Hamiltonian \eqref{H} can be
rewritten as
\begin{equation}\label{03}
\hat{{\cal H}}(t)= \hat{{\cal H}}_{0}+\left[\hat{V}_1e^{i\omega
t}+\hat{V }_2e^{i2\omega t}+\mathrm{H.c.}\right],
\end{equation}
where the time-independent part is
\begin{equation}\label{H0}
\hat{\cal H}_{\mathrm{0}} = \hat{\mathcal{H}}_{\mathrm{L}} +
\hat{\mathcal{H}}_{\mathrm{BIA}}
+\left(\gamma_1+{5\gamma_2}/{2}\right) {k}^2_0
-\gamma_2(J_x^2+J_y^2){k}^2_0,
\end{equation}
and the two harmonics are
\begin{eqnarray}\label{V1}
\hat{V}_1&=&[(\gamma_1+{5}\gamma_2/2)
(k_x-ik_y)-2\gamma_2(J_x^2k_x-iJ_y^2k_y)\nonumber\\
&-&\gamma_3(\{J_x, J_y\}(k_y-ik_x)+\{J_x-iJ_y, J_z\}k_z)\nonumber\\
&+&({\alpha}/{2})(\{J_x, (J_y^2-J_z^2)\}-i\{J_y,
(J_z^2-J_x^2)\})]k_0,\nonumber\\
\hat{V}_2&=&[({\gamma_2}/{2})(J_y^2-J_x^2)+i(\gamma_3/2)\{J_x,
J_y\}]{k}^2_0.
\end{eqnarray}
Here, $k_0=|e|E/\hbar\omega$ is the amplitude of the field-induced
shift of the in-plane electron wave vector. Applying the
conventional Floquet-Magnus approach~\cite{FM1,FM2,FM3} to
renormalize the time-dependent Hamiltonian (\ref{03}), we arrive
at the effective time-independent Hamiltonian,
\begin{align}\label{12}
&\hat{{\cal H}}_{\mathrm{eff}}=\hat{{\cal
H}}_0+\frac{\left[\hat{V}_1,\hat{V}^\dagger_{1}\right]}{\hbar\omega}+\frac{\left[\hat{V}_2,\hat{V}^\dagger_{2}\right]}{2\hbar
\omega}\nonumber\\
&+\frac{\left[\left[\hat{V}_1,\hat{{\cal
H}}_0\right],\hat{V}^\dagger_{1}\right]+\mathrm{H.c.}}{2(\hbar\omega)^2}+\frac{
\left[\left[\hat{V}_2,\hat{{\cal H}}_0\right],\hat{V}^\dagger_{2}\right]+\mathrm{H.c.}}{8(\hbar\omega)^2}\nonumber\\
&+\mathrm{\it o}\left(\frac{\hat{V}_{1,2}}{\hbar\omega}\right)^2,
\end{align}
where the square brackets $[\hat{A},\hat{B}]$ represent the
commutators of the operators $\hat{A}$ and $\hat{B}$. In what
follows, we consider the dressing field (\ref{A}) to be
high-frequency (${\gamma_{1,2,3}k_0^2}/{\hbar\omega}\ll1$ and
${\alpha k_0}/{\hbar\omega}\ll1$). For reasonable irradiation
intensities of kW/cm$^2$ scale, these conditions can be satisfied
for high-frequencies beginning from the THz range. In this
high-frequency limit, the effective time-independent Hamiltonian
(\ref{12}) reads $\hat{{\cal H}}_{\mathrm{eff}}\approx\hat{{\cal
H}}_0$. Therefore, one can use the stationary Hamiltonian
(\ref{H0}) to describe the electronic states near the $(001)$
surface renormalized by the high-frequency field (\ref{A}). In the
Luttinger-Kohn basis, $\psi_{j_z}$, the Hamiltonian (\ref{H0})
reads
\begin{equation}\label{LHM}
\hat{\cal H}_0 =
\begin{tabular}{|c||c c c c|} \hline
${j_z}\backslash {j_z}$ & ${+3/2}$ & ${-1/2}$ & ${+1/2}$ & ${-3/2}$ \\
\hline\hline ${+3/2}$ & $F_0$ & $I+L$ & $H+M$ & $N$ \\
${-1/2}$ & $I^*+L$ & $G_0$ & $-N$ & $-H+M$ \\
${+1/2}$ & $H^*+M^*$ & $-N^*$ & $G_0$ & $I-L$ \\
${-3/2}$ & $N^*$ & $-H^*+M^*$ & $I^*-L$ & $F_0$ \\
\hline
\end{tabular}\,,
\end{equation}
where the field-containing matrix elements are
$F_0=F+\Delta_0+\Delta/2$ and $G_0=G+\Delta_0-\Delta/2$,
$\Delta=2\gamma_2(eE/\hbar\omega)^2$ is the field-induced
splitting of the conduction and valence bands in the Brillouin
zone center, and $\Delta_0=\gamma_1(eE/\hbar\omega)^2$ is the
field-induced shift of zero energy, which is omitted in the
following.

\section{Results and Discussion}

First of all, let us consider the electronic states with the zero
in-plane wave vector, $k_x=k_y=0$. For these states, the Floquet
engineered Hamiltonian (\ref{LHM}) takes the block-diagonal form
\begin{equation}\label{Hk0}
\hat{\cal H}_0 =
\begin{bmatrix}
\,\,\hat{\cal H}^{\pm}\,\, & \,\,0\,\,\\
\,\,0\,\,&\,\,\hat{\cal H}^{\mp}\,\,
\end{bmatrix},
\end{equation}
where
\begin{equation}\label{Hpm}
\hat{\cal H}^{\pm}=
\begin{bmatrix}
\,\,(\gamma_1\mp2\gamma_2){k}_z^2\pm\Delta/2\,\,&\,\,\pm\sqrt{3}\alpha{k}_z\,\,\\
\,\,\pm\sqrt{3}\alpha{k}_z\,\,&\,\,(\gamma_1\pm2\gamma_2){k}_z^2\mp\Delta/2\,\,
\end{bmatrix}.
\end{equation}
The four eigenspinors of the Hamiltonian (\ref{Hk0}) describing
bulk electronic states of HgTe at $k_x=k_y=0$ can be written as
\begin{equation}\label{Phi}
\varphi_j^+=
\begin{bmatrix}
\lambda_j^+(k_z)\\
1\\
0\\
0
\end{bmatrix}e^{ik_zz},\,\,\varphi_j^-=
\begin{bmatrix}
0\\
0\\
1\\
\lambda_j^-(k_z)
\end{bmatrix}e^{ik_zz},\,\,j=1,2,
\end{equation}
where
\begin{equation}\label{lambda}
\lambda_j^{\pm}(k_z)=\frac{\mp\sqrt{3}\alpha
k_z}{(\gamma_1\mp2\gamma_2)k_z^2\pm\Delta/2-\varepsilon_j(k_z)},
\end{equation}
and the corresponding eigenenergies are
\begin{equation}\label{eps}
\varepsilon_j(k_z)=\gamma_1k_z^2+(-1)^j\sqrt{(\Delta/2-2\gamma_2k_z^2)^2+3\alpha^2k_z^2}.
\end{equation}
The electronic states localized near the surface $(001)$ are
described by the same spinors (\ref{Phi}) with the imaginary
$z$-component of electron wave vector, $k_z=i\kappa$. The two
energy branches (\ref{eps}) produce two different parameters
$\kappa=\kappa_{1,2}$ satisfying the equation
\begin{equation}\label{eps0}
\varepsilon=\gamma_1\kappa_j^2+(-1)^j\sqrt{(\Delta/2+2\gamma_2\kappa_j^2)^2-3\alpha^2\kappa_j^2},
\end{equation}
where $\varepsilon$ is the energy of the surface state. It should
be stressed that the parameters $\kappa_{1,2}$ can be complex
numbers but their real part must be positive for spinors
(\ref{Phi}) to decay exponentially into the bulk at
$z\rightarrow\infty$ (see Fig.~\ref{Fig1}b). Making the
replacement, $k_z\rightarrow i\kappa_{1,2}$, in
Eqs.~(\ref{Phi})-(\ref{eps}), one can write the surface-localized
eigenfunction of the Hamiltonian (\ref{Hk0}) as a linear
combination of the spinors (\ref{Phi}),
\begin{align}\label{psi}
&\Psi(z)=C_1\begin{bmatrix}
\lambda_1^+(i\kappa_1)\\
1\\
0\\
0
\end{bmatrix}e^{-\kappa_1z}+C_2\begin{bmatrix}
\lambda_2^+(i\kappa_2)\\
1\\
0\\
0
\end{bmatrix}e^{-\kappa_2z}\nonumber\\
&+C_3\begin{bmatrix}
0\\
0\\
1\\
\lambda_1^-(i\kappa_1)
\end{bmatrix}e^{-\kappa_1z}+C_4\begin{bmatrix}
0\\
0\\
1\\
\lambda_2^-(i\kappa_2)
\end{bmatrix}e^{-\kappa_2z},
\end{align}
where $C_{1,2,3,4}$ are the constants to be determined. To do so,
we chose the model of a surface potential which can be
approximated by the infinitely-high barrier at the coordinate
$z=0$. This sets the boundary condition for the electron wave
function (\ref{psi}) as $\Psi|_{z=0}=0$, and results into a
homogeneous system of four algebraic equations defining the
constants $C_{1,2,3,4}$,
\begin{eqnarray}\label{C}
C_1\lambda_1^+(i\kappa_1)+C_2\lambda_2^+(i\kappa_2)&=&0,\,\,\,
C_1+C_2=0,\nonumber\\
C_3\lambda_1^-(i\kappa_1)+C_4\lambda_2^-(i\kappa_2)&=&0,\,\,\,C_3+C_4=0.
\end{eqnarray}
The secular equation for the algebraic system (\ref{C}),
\begin{equation}\label{dett}
[\lambda_1^+(i\kappa_1)-\lambda_2^+(i\kappa_2][\lambda_1^-(i\kappa_1)-\lambda_2^-(i\kappa_2)]=0,
\end{equation}
defines the sought energy (\ref{eps0}) of the surface electronic
states at $k_x=k_y=0$.

The system of equations (\ref{dett}) and (\ref{eps0}) can be
easily solved analytically if $\gamma_1=0$. Physically, this
particular case corresponds to a semiconductor with the
Hamiltonian (\ref{HL}), where the masses of electrons and holes
along the $z$ axis, $m_e$ and $m_h$, are equal to each other. For
such a symmetric electron-hole system, the eigenenergy
(\ref{eps0}) is $\varepsilon=0$ and the surface-localized
eigenspinors (\ref{psi}) corresponding to this eigenenergy can be
written as
\begin{eqnarray}
\Psi_1(z)&=&A(e^{-\kappa_1z}-e^{-\kappa_2z})\,\begin{bmatrix}
-i\\
1\\
0\\
0
\end{bmatrix},\label{Psi1}\\
\Psi_2(z)&=&A(e^{-\kappa_1z}-e^{-\kappa_2z})\,\begin{bmatrix}
0\\
0\\
1\\
i
\end{bmatrix},\label{Psi2}
\end{eqnarray}
where
\begin{equation}\label{kappa}
\kappa_j=\left|\frac{\sqrt{3}\alpha}{4\gamma_2}\right|+(-1)^j\sqrt{\left(\frac{\sqrt{3}\alpha}{4\gamma_2}\right)^2-\frac{\Delta}{4\gamma_2}},
\end{equation}
and $A=\sqrt{\sqrt{3}\alpha\Delta/(6\alpha^2-8\gamma_2\Delta)}$ is
the normalization constant. The eigenspinors
(\ref{Psi1})-(\ref{Psi2}) can be easily verified by direct
substitution into the Schr\"odinger equation, $\hat{\cal
H}\Psi_{1,2}=\varepsilon\Psi_{1,2}$, with the Hamiltonian
(\ref{Hk0}) and the eigenenergy $\varepsilon=0$. Since
$\kappa_{1,2}\neq0$ if $\alpha\neq0$, the surface states
(\ref{Psi1})-(\ref{Psi2}) physically originate from the BIA of the
crystal. To find the dispersion of the surface states
(\ref{Psi1})-(\ref{Psi2}) for small in-plane wave vectors
${k}_{x,y}$, we have to project the total Hamiltonian (\ref{LHM})
to the subspace spanned by these two states, $\Psi_1$ and
$\Psi_2$. Keeping the terms linear in ${k}_{x,y}$, we arrive at
the Hamiltonian,
\begin{equation}\label{Heff}
\hat{\cal
H}_{\mathrm{D}}=-\frac{3\alpha}{2}(\sigma_xk_x+\sigma_yk_y)-\frac{\sqrt{3}\alpha}{2}
(\sigma_xk_y+\sigma_yk_x),
\end{equation}
where $\sigma_{x,y}$ are the Pauli matrices written in the basis
(\ref{Psi1})-(\ref{Psi2}). Diagonalizing the Hamiltonian
(\ref{Heff}), we can write the sought energy spectrum of the
surface states (\ref{Psi1})-(\ref{Psi2}) near $k_x=k_y=0$ as
\begin{equation}\label{ED}
\varepsilon(k_x,k_y)=\pm\sqrt{3}\alpha\sqrt{k_x^2+k_y^2+\sqrt{3}k_xk_y}.
\end{equation}
Eqs.~(\ref{Heff})-(\ref{ED}) reveal the energy spectrum of the
found surface states, which is typical for topological
insulators~\cite{Hasan_2010}. Namely, the two degenerate states
(\ref{Psi1})-(\ref{Psi2}) form the Dirac point at ${k}_x=k_y=0$
with the energy $\varepsilon=0$, which lies in the middle of the
conduction and valence bands, and the linear dispersion (\ref{ED})
near the point appears.
\begin{figure}[h]
\includegraphics[width=1\linewidth]{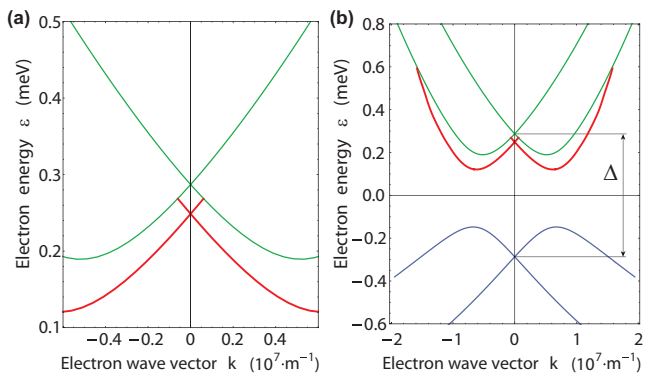}
\caption{The energy spectrum of the optically induced topological
electronic states (bold red curves) on the $(001)$ surface of HgTe
irradiated by a circularly polarized electromagnetic wave with the
intensity $I=850$~W/cm$^2$ and the photon energy
$\hbar\omega=2$~meV: (a) near the Dirac point; (b) in the broad
range of electron wave vectors. Thin green and blue curves
represent the dispersion of conduction and valence bands in bulk
HgTe, respectively, and $\Delta$ is the optically induced band
gap.} \label{Fig2}
\end{figure}

The approach discussed above, which describes analytically the
surface states originated from the Hamiltonian (\ref{LHM}) at
$k_x=k_y=0$ and $\gamma_1=0$, can be applied numerically to the
same Hamiltonian (\ref{LHM}) at any wave vectors $k_{x,y}$ and
parameters $\gamma_{1,2,3}$ as well. As a result, one can
calculate the energy spectrum of the surface states in the broad
range of electron wave vectors and band parameters. The
numerically calculated dispersion of the surface states is
presented in Fig.~2 for the following HgTe band
parameters~\cite{Adachi_2004,Ruan_2016}:
$\gamma_1=15.6\,\hbar^2/2m_0$, $\gamma_2=9.6\,\hbar^2/2m_0$,
$\gamma_3=8.6\,\hbar^2/2m_0$ and $\alpha=0.208$~\AA$\cdot$eV.
Since the electron-hole system in HgTe is strongly asymmetric,
$m_e/m_h=(\gamma_1-2\gamma_2)/(\gamma_1+2\gamma_2)\ll1$, the Dirac
point energy is shifted towards the conduction band (see Fig.~2a).
It is seen also that the branches of the surface states merge into
the spectrum of bulk conduction band if the plane electron wave
vector, ${k}_{x,y}$, is large enough (see Fig.~\ref{Fig2}b). As a
result, the discussed surface states as a whole are localized near
the conduction band of HgTe for small wave vectors $k$. It should
be stressed that the Hamiltonian (\ref{Heff}) and the dispersion
(\ref{ED}) are applicable to describe the energy spectrum of
surface states near the Dirac point for any band parameters.
Particularly, the Dirac velocity, $v_D=\sqrt{3}\alpha/\hbar$,
which can be extracted from the dispersion (\ref{ED}), does not
depend on the Luttinger parameters $\gamma_{1,2,3}$. We note that
the effective Hamiltonian (\ref{LHM}) is similar to the
Hamiltonian of a strained gapless semiconductor~\cite{Bir_Pikus}.
Therefore, the discussed topological states behave like those in
strained HgTe~\cite{Ruan_2016}.
\begin{figure}[h]
\includegraphics[width=1\linewidth]{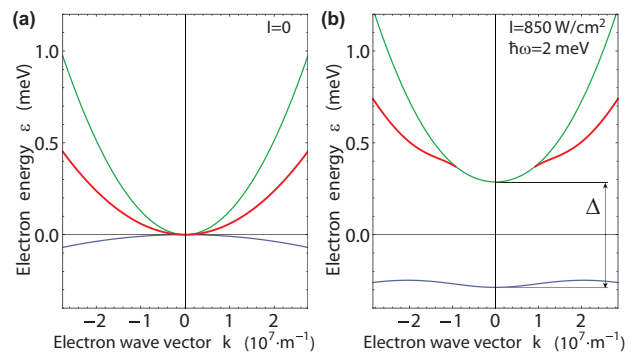}
\caption{The energy spectrum of the D'yakonov-Khaetskii surface
electronic states (bold red curves) on the $(001)$ surface of
HgTe:(a) in the absence of irradiation; (b) in the presence of a
circularly polarized electromagnetic wave with the intensity
$I=850$~W/cm$^2$ and the photon energy $\hbar\omega=2$~meV. Thin
green and blue curves represent the dispersion of bulk bands,
respectively, and $\Delta$ is the optically induced band gap.}
\label{Fig3}
\end{figure}

It follows from Eq.~(\ref{kappa}) that $\kappa_1\neq0$ if the
optically induced band gap is $\Delta\neq0$. Otherwise,
$\kappa_1=0$ and the corresponding electronic wave function
plotted in Fig.~1b is delocalized, i.e. the surface states vanish.
Therefore, the surface electronic states discussed above exist
only in the presence of the irradiation and, thus, are optically
induced. This is the substantial difference of the considered
surface states from the non-topological surface states existing in
gapless HgTe without an irradiation, which were analyzed for the
first time by D'yakonov and Khaetskii~\cite{Dyakonov_1981}. To
demonstrate the difference between these two kinds of surface
states in more details, let us consider the evolution of the
D'yakonov-Khaetskii (DKh) surface states under the irradiation. In
contrast to the considered optically induced surface states, the
BIA Hamiltonian (\ref{Hbia}) is not crucial for their existence.
Therefore, we will assume $\alpha=0$ in the following analysis. To
simplify calculations, we also neglect the weak anisotropy of
electron-hole dispersion in HgTe. Mathematically, this corresponds
to the replacement of the Luttinger parameters,
$\gamma_{2,3}\rightarrow\gamma=(2\gamma_2+3\gamma_3)/5$, in the
Hamiltonian (\ref{LHM}). Under the made assumptions, the
surface-localized eigenspinor of the Hamiltonian (\ref{LHM}) can
be written as
\begin{align}\label{psiDKh}
&\Psi(z)=C_1\begin{bmatrix}
e^{-\frac{i3\theta}{2}}\\
\lambda_1^+e^\frac{i\theta}{2}\\
-i\lambda_1^+e^{-\frac{i\theta}{2}}\\
-ie^{\frac{i3\theta}{2}}
\end{bmatrix}e^{-\kappa_1z}+C_2\begin{bmatrix}
e^{-\frac{i3\theta}{2}}\\
\lambda_1^-e^\frac{i\theta}{2}\\
i\lambda_1^-e^{-\frac{i\theta}{2}}\\
ie^{\frac{i3\theta}{2}}
\end{bmatrix}e^{-\kappa_1z}\nonumber\\
&+C_3\begin{bmatrix}
i\lambda_2^+e^{-\frac{i3\theta}{2}}\\
-ie^\frac{i\theta}{2}\\
e^{-\frac{i\theta}{2}}\\
-\lambda_2^+e^\frac{i3\theta}{2}
\end{bmatrix}e^{-\kappa_2z}+C_4\begin{bmatrix}
-i\lambda_2^-e^{-\frac{i3\theta}{2}}\\
ie^\frac{i\theta}{2}\\
e^{-\frac{i\theta}{2}}\\
-\lambda_2^-e^\frac{i3\theta}{2}
\end{bmatrix}e^{-\kappa_2z},
\end{align}
where
\begin{equation}\label{lDKh}
\lambda_{j}^\pm=\frac{(-1)^j[\varepsilon-\gamma_1(k^2-\kappa_j^2)]+\gamma(k^2+2\kappa_j^2)
+\Delta/2}{\sqrt{3}\gamma(k^2\pm2\kappa_jk)},
\end{equation}
$\mathbf{k}=(k_x,k_y,0)=(k\cos\theta,\,k\sin\theta,\,0)$ is the
in-plane wave vector, and the energy of the surface electronic
states is
\begin{align}\label{energyDKh}
&\varepsilon=\gamma_1(k^2-\kappa_j^2)\nonumber\\
&+(-1)^j\gamma\sqrt{3k^4-12\kappa_j^2k^2+[k^2+2\kappa_j^2+\Delta/(2\gamma)]^2}.
\end{align}
Applying the zero boundary condition, $\Psi(0)=0$, to the
eigenspinor (\ref{psiDKh}), we arrive at the homogeneous system of
four algebraic equations defining the constants $C_{1,2,3,4}$. The
secular equation of the system reads as
\begin{equation}\label{dettDKh}
[\lambda_1^+\lambda_2^-+1][\lambda_1^-\lambda_2^++1]=0.
\end{equation}
Solving this secular equation, one can find the energy spectrum of
the DKh surface states in irradiated HgTe, $\varepsilon$. In the
absence of the irradiation ($\Delta=0$), Eq.~(\ref{dettDKh}) can
be solved analytically and leads to the known dispersion of the
DKh surface states~\cite{Dyakonov_1981},
\begin{equation}\label{DKh}
\varepsilon=
\left[1-\left(\frac{1+\sqrt{3(2\gamma-\gamma_1)/(2\gamma+\gamma_1)}}{2}\right)^2\right]
(\gamma_1+2\gamma){k^2},
\end{equation}
which is plotted in Fig.~\ref{Fig3}a. Since the DKh surface states
are not topological, their dispersion is parabolic in contrast to
the Dirac dispersion of the optically induced states plotted in
Fig.~2. Solving Eq.~(\ref{dettDKh}) numerically for $\Delta\neq0$,
we arrive at the spectrum of the DKh states on the irradiated
surface, which is shown in Fig.~\ref{Fig3}b. We observe that the
discussed states exist only for large electron wave vectors, $k$,
and vanish near $k=0$. Namely, it follows from Fig.~\ref{Fig3}b
that the branch of the DKh surface states merges into the
continuum of bulk conduction band at a some critical electron wave
vector. The value of the critical wave vector, $k=k^{\,\prime}$,
is defined by Eq.~(\ref{dettDKh}), where the energy of the surface
electron states (\ref{energyDKh}) is equal to the energy of bulk
conduction band in irradiated HgTe. Taking this into account, one
can find $k^{\,\prime}\propto\sqrt{\Delta}$. Thus, the irradiation
shifts the existence domain of the DKh states to the region of
large electron wave vectors, $k$, inside the conduction band. As a
consequence, they disappear near the Dirac point of the optically
induced topological states shown in Fig.~\ref{Fig2}a. It should be
noted also that the DKh states do not lie within the band gap
$\Delta$ (see Fig.3b). Therefore, they cannot turn mercury
telluride into topological insulator.

\section{Conclusion}
We developed the theory which describe surface electronic states
appearing on the surface of HgTe due to the mixing of the
conduction and valence bands by a circularly polarized
electromagnetic field. The states originate from the bulk
inversion asymmetry of HgTe and have the Dirac point in their
dispersion, which is characteristic for topological states. It is
shown that the structure of these optically induced topological
states differs substantially from the known non-topological
D'yakonov-Khaetskii surface states~\cite{Dyakonov_1981} existing
in HgTe in the absence of irradiation. Namely, the irradiation
shifts these surface states to the region of large electron wave
vectors, whereas the optically induced topological states are
localized near the Brillouin zone center. As a consequence, these
two kinds of surface states can be detected in experiments
independently. It should be noted that the experimental
methodology based on the angle resolved photoemission spectroscopy
(ARPES) technique, which is commonly used to study surface
electronic states in various condensed-matter
structures~\cite{Chen_2012_1,Wang2013,Mahmood2016}, is also
appropriate for observation of the optically induced topological
states discussed above. Since the energy difference between the
states and the conduction band is of sub-meV scale (see Fig.~2),
the temperatures around 1K are required to observe them
experimentally.

\begin{acknowledgements}
The work was partially supported by Horizon2020 RISE project
COEXAN, Rannis project 163082-051, Russian Foundation for Basic
Research (project 17-02-00053), Ministry of Education and Science
of Russian Federation (projects 3.4573.2017/6.7, 3.2614.2017/4.6,
14.Y26.31.0015), and the Government of the Russian Federation
through the ITMO Fellowship and Professorship Program. O.K. and
O.V.K. thank the University of Iceland for hospitality.
\end{acknowledgements}

\end{document}